\documentclass[onecolumn,a4paper,9pt]{article}
\usepackage[utf8]{inputenc}
\usepackage[T1]{fontenc}
\usepackage{lmodern,textcomp}
\usepackage[english]{babel}
\usepackage[top=2.5cm,bottom=2.5cm,left=2.5cm,right=2.5cm,headsep=1cm]{geometry}
\usepackage{amsfonts,amsmath,amssymb}
\usepackage{color}
\usepackage{graphicx}
\usepackage{caption}
\usepackage{subfigure}
\usepackage{array}
\usepackage{multirow}
\usepackage{pict2e}
\usepackage{fancyhdr}
\title{Speckle characterization in a cinematography projection configuration}

\author{Pierre Walczak$^{1,*}$,Xavier Hachair$^{2}$,Stéphane Barland$^{1}$}

\date{%
\small{$^{1}$Universit\'e Côte d'Azur, CNRS, Institut de Physique de Nice, Sophia Antipolis, France\\%
$^{2}$BBright, Les Galaxies, 1st floor, 1 Square René Cassin, Rennes,France}\\%
$^{*}$Corresponding author: pierre.walczak@inphyni.cnrs.fr\\%
\today}

\begin{document}

\maketitle
\thispagestyle{fancy}

\textbf{Due to high exploitation costs and other environmental issues, it would be desirable to phase out large cinema projection systems based on standard xenon lamps in favor of laser based projection devices. Lasers provide longer lifetime, wider color gamut, smaller extension and higher stability of light output. But the high degree of coherence of theses sources also lead to the formation of granular structures, usually known as speckle. 
When an imaging system is involved, as in the cinema projection case because of the capacity of the human eye to form an image of the screen, we speak about subjective speckle. In order to remove this spatial random pattern, different methods have been studied as temporal and/or spatial coherence reduction. But most of them can’t be used in the context of cinema projection because they don’t respect the cinematography projection standard. In our work, we have studied the possibility to reduce the subjective speckle either by changing the coherence of the light source or by studying the influence of the different elements constituting the projection display and most importantly in the conditions imposed by cinematography industry. Thanks to a lasers array formed by N independant semiconductor lasers, we have measured the evolution of the subjective speckle contrast in function of the number of sources. The resulting contrast discreases as a square root function and reach a saturation level when a light pipe is used. This behavior is directly due to the caracteristic of the light pipe which limit at its output the spatial coherence of the light source. Futhermore, in a different configuration, we have studied the influence of diffusers and the magnification of the projector zoom. It has been demonstrated that magnification plays an important rôle on the speckle formation because it increases the coherence length determined by the light pipe. On the contrary, the diffusers placed before the light pipe doesn’t change really the amount of subjective speckle.}
\vspace{1cm}

\section*{Introduction}
The light source at the origine of the cinema projection consists to use discharge lamps as for exemple carbon arc bulbs for the first one and since 1940's the Xenon lamps. The characteristics of those light sources make the energy efficiency of projection systems low mainly because of the decomposition-recomposition to the Red, Green, Blue (RGB) colors and the polarization effect. In addition these sources create images with low color saturation (bad color gamut). These last decades, the evolution of laser technology is found to be a nice successor to the standard light sources based on filament lamps. Lasers have the advantage to be monochromatic, directional, linear polarized and coherent. The projection system based on RGB lasers sources provide a higher color gamut, a higher energy efficiency and a better life time. However, the main problem of these sources is to create the well-known phenomenon called speckle due to the coherence of the lasers.   

Speckle is unavoidably generated by interference when a scattered highly coherent wave, such as lasers, is reflected by a random surface roughness, such as typical cinema screens. It is common to differentiate objective and subjective speckle. Speckle patterns are called objective when any imaging system is involved. The speckle distribution can then be measured directly on an imaging sensor without optical lens. On the contrary, when an imaging system is involved to form the image of the screen on a detector (camera or human retina), we speak about subjective speckle. It doesn't exist a standardized quantification of speckle patterns. The more using tool to quantify speckle is to calculate its contrast value defined by the ratio of the standard deviation to the mean of the light intensity fluctuations. When the contrast is unity, the fluctuations of intensity are strong and the probability density function is exponential. The speckle is generally called as fully developped speckle~\cite{Goodman:2007} and it will be generated when the surface roughness is comparable with or larger than the wavelength of the light source. On the opposite, when the contrast is null, the fluctuations are very small and the speckle doesn't exit no longer.

The theory of speckle formation, its statistical properties, and methods of speckle reduction have been studied thoroughly those last decades~\cite{Goodman:2007,Goodman:2015}. Several schemes for the reduction of speckles have been developped such as polarization diversity~\cite{Goodman:2007}, wavelength and/or angular diversity~\cite{Goodman:2007,George:1973}. But the most common method is a moving diffuser placed in the path of a laser beam. In this configuration, the phase contribution in every point of the image is modified in function of time. It is possible to use one or several diffusers~\cite{Lowenthal:1971}, to introduce a vibration~\cite{Kubota:2010,Yao:2012,Pan:2014} or a rotation~\cite{Kubota:2010} of the diffusing element. Moreover, many other methods exist, such as using a laser array~\cite{Zhang:2012}, moving the observation screen~\cite{Rawson:1976}, scanning micromirrors~\cite{Yurlov:2008,Akram:2010}, diffractive optical elements~\cite{Wang:1998,Ouyang:2010,Yao:2012}, rotating light pipe~\cite{Sun:2010} or vibrating multimode optical bundle~\cite{Petoukhova:2004,Mehta:2012}. Howevever, few of these methods can be applied in the specific context of cinema laser projection. Indeed, this context introduces specific normative and technological constraints. From the normative point of view, the light emitted by the projection system must satisfy several criteria imposed by the Digital Cinema Initiatives (DCI), as for instance a typical value for the luminance. From the technological point of view, the projection system must include a homogeneizing element, whose image will be formed on an intensity modulation device (Digital Light Processor, CMOS, ...) before being finally projected onto the screen. Several types of homogeneizing devices can be used (light pipe~\cite{Roelandt:13,Pan:2014}, bi-convex aspheric lens~\cite{zhang:2007}, or lenslet integrator~\cite{Yao:2012} as double-sided microlens array~\cite{Chen:2009}) but in any case the zooming system following this device will finally project its output on the screen, after application of a spatial intensity modulation corresponding to the scene content. This point is key, as it implies that any speckle reduction device altering the spatial intensity or phase profile of the light must be placed upstream from the homogeneizing device if optimal image quality is to be preserved. Here we adress the formation of subjective speckle in model experiments which feature the key elements of video projection devices. Specifically, we analyze the impact of the spatial coherence of the light source in absence and presence of the required homogenizing elements (first section), the impact of diffusors upstream the light pipe (second section) and that of the zooming system leading to suitable size of image projection (third section).

\section*{Speckle modification by the source coherence control}

In the first part of our study, we are concentrated on the laser source light and its spatial coherence properties on the formation of the speckle. Then, we have placed as it is recommended for the cinema projection condition a diffuser and a light pipe to study their impact on the speckle. In this section, we are not in the standard illumination recommended by DCI but it is the only exception of our study.

\begin{figure}[h!]
 \centering
\includegraphics[scale=0.4]{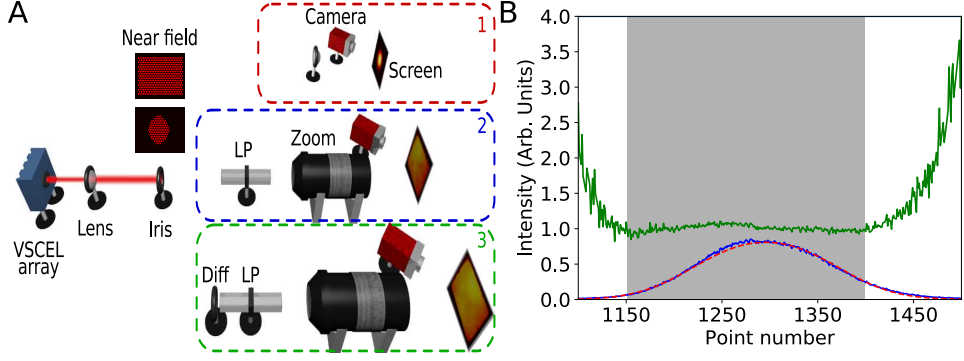}
\caption{{\bf Illustration of the experimental setup and cross section of the far field experiment.}
(A) We have represented the experimental setup. It is subdivided in three parts corresponding to three different experiments. We have measured subjective speckle directly formed by the light source (configuration 1). Then, we have investigated the influence of the light pipe only (configuration 2). Finally, we have introduced different diffusers (configuration 3). (B) A cross section section of the far field is traced in blue line, the 2D fit in red dashed line and the resulting speckle in green line. The grey area illustrate area where the speckle is calculated.}
\label{fig1}
\end{figure}

Fig.~\ref{fig1}A shows a schematic representation of the experimental setup inplemented to study the influence of the light source on the subjective speckle formation. The first part of the setup is common to all our experiments detailled in this section. The light source is a VCSEL array with a thermal regulation and a laser diode controller. The VCSEL array (Philips PLA1520-980) is a small-sized, infrared laser array light source based on Philips proprietary Vertical Cavity Surface Emitting Laser (VCSEL) technology. It consists of over 450 VCSELs emitting at 980nm with a typical power of 2W and a spectral width of 2nm. After the source, the near field of the VCSEL array is formed thanks to a lens. At the near field plane, an iris diaphgram allows to select the number of sources used to make the experiment. In this way, we are able to select between 1 to 450 lasers as we can see on the Fig.~\ref{fig1}A. So, we can analyse the subjective speckle formed by the sum of N independant coherent sources where we can control the parameter N. After the iris, we have performed three different experiments that we have known configuration 1, 2 and 3 in the Fig~\ref{fig1}A.

\begin{figure}[h!]
   \centering
\includegraphics[scale=0.65]{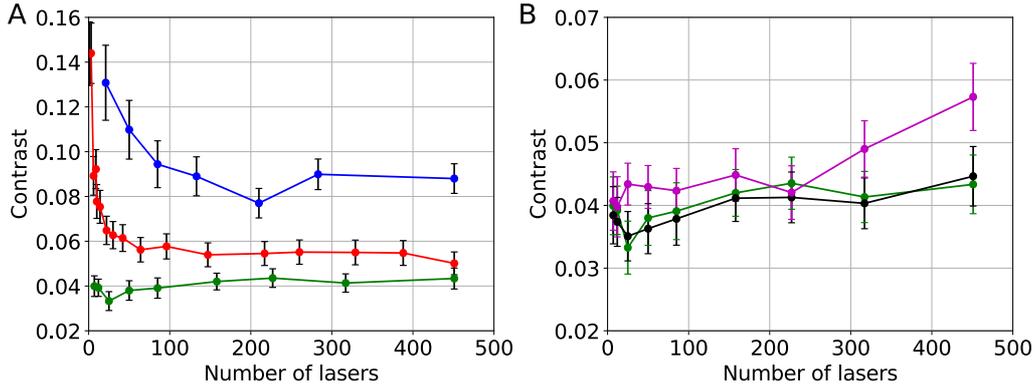}
\caption{{\bf Contrast variations in function of the source coherence.}
A: We have calculated the speckle contrast for three different experiments. We have analysed the impact of source light thanks to the far field, the influence of the light pipe and the effect of diffuser + light pipe, respectively in red, blue and green line. B: We are in a configuration where the light pipe and the diffuser are used. We have studied the influence of the diffuser granularities (grit) on the speckle. The green, black and magenta line correspond respectively to a diffuser of 120grit, 220grit and 600grit.}
\label{fig2}
\end{figure}

In the first configuration of the Fig.~\ref{fig1}A, we have investigated directly the importance of the light source on the speckle formation by projected the far field of the beam after the iris on a paper screen. After reflection, we have recorded the subjective speckle on a camera (Allied Vision Mako G-419B NIR). At typical operating current for all the lasers, the far field emission pattern is donut shape. We have represented on the Fig.~\ref{fig1}B a cross section of the far field in blue line. His shape is close to a gaussian function. In order to estimate the speckle contrast, we have used the following procedure. We have performed a higher order gaussian fit where the content of the exponent is a fit parameter. We have obtained the red-dashed line on the Fig.~\ref{fig1}B. These two curves are almost similar. In order to extract the speckle fluctuations localized around the maximum of the gaussian data we have divided the data curve by the fit curve. The results of this operation is illustrated by the green curve on the Fig.~\ref{fig1}B. The speckle contrast (C) is calculated inside the grey surface area from the following well-known equation:
\begin{eqnarray}
\label{eq:contrast}
C=\frac{\sigma}{\mu}
\end{eqnarray}
where $\mu$ is the mean value, $\sigma$ the standard deviation :
\begin{eqnarray}
\label{eq:kurtMoyStd}
\mu=\frac{1}{n}\sum_{i=1}^{n} X_{i};\sigma=\left[\frac{1}{n-1}\sum_{i=1}^{n}(X_{i}-\mu)^{2}\right]^{1/2}
\end{eqnarray}
As mentioned earlier, we can vary the number of lasers and measure the corresponding subjective speckle. The result is represented in the red line on the Fig.~\ref{fig2}A. We obtain an exponential decrease from the speckle formed by one laser with a contrast of 14.4\% to the speckle formed by 450 lasers with a contrast of 5\%. This strong decrease of contrast has been theoritically demonstrated by Goodman~\cite{Goodman:2007}. Our VCSEL array can be considerated as a sum of N independant speckle intensities with the same mean power value. The theory predict that the contrast falls in proportion to $1/\sqrt{N}$ as the independent patterns N increases. But in our experiment we don't reach the maximum value contrast when we have only one laser. We attribute this value, first, to the presence of noise, and to the finite laser linewidth.  

In order to improve our contrast estimation, we have calculated a theoritical error $\delta$ for each measure, which is represented by the errorbar on the Fig~\ref{fig2}A,B and it corresponds to the following formula:
\begin{eqnarray}
\label{eq:delta}
\delta=C N^{-1/4} \left(k-\frac{N-3}{N-1}\right)^{1/4}
\end{eqnarray}
where kurtosis $k$ of the intensity distribution is calculated from the Eq~\ref{eq:kurtosis}.
\begin{eqnarray}
\label{eq:kurtosis}
k=\frac{1}{n-1}\sum_{i=1}^{n} \left(\frac{X_{i}-\mu}{\sigma}\right)^{4}
\end{eqnarray}

We have measured the speckle with only the influence of the light source. In the second and third configuration illustrated on the Fig~\ref{fig1}A, we have studied the role of the light pipe only (configuration 2 in blue) and then the impact of the diffuser coupled with the light pipe (configuration 3 in green). In these configurations, we have projected the output face of the lightpipe on the same paper screen thanks to a projector zoom (minolta DLP projector zoom XGA) typically used in the cinema projection display. So, the field at the output of the iris diaphragm propagates through all these optical element to the screen and the subjective speckle is recorded on the camera.

The contrast calculated in the second configuration where we have used only the lightpipe is plotted in blue line on the Fig~\ref{fig2}A. A exponential decrease is also observed but we reach a saturation level with a contrast value of 8.8\%. The light pipe is identical to a frequential filter. If only a part of the lasers source are porjected on the spatial modes of the light pipe, then only these modes are homogenized. The other pass trough the light pipe without modification. So, on the screen we are not a complete superposition of all the lasers. 

Finally, in the third configuration with the diffuser and the light pipe, in green line on the Fig~\ref{fig2}A, the contrast doesn't really evolve and follow a constant evolution if we take account of our theoritical error. In this case, we have used a ground glass diffuser of 120grit (package Thorlabs DGK01). In function of the grid, the diffuser has a surface more or less rough. By definition, a finer grain (higher number) allows a better transmission efficiency. The results for the coaser diffuser (120grit) is illustrated in green line on the Fig.~\ref{fig2}A and~\ref{fig2}B and respectively in black line and magenta line for the 220grit and 600grit diffuser. The results for the 120grit and the 220grit are very similar in the measure incertainty. But for the 600grit diffuser, the contrast is a little bit higher especially for the higher number of lasers.

\section*{Influence of the diffusers on the subjective speckle}
In the last section, we have studied the influence of the source coherence on the formation of subjective speckle. But as we have mentionned in the introduction, with a power of 2W, we are not in the cinema projection standard. For this reason, we have changed the source and we have projected the output of the light pipe on a typical cinema projection screen in order to preserve the standard luminance which is about 12cd/m$^{2}$.

\begin{figure}[h!]
   \centering
\includegraphics[scale=0.5]{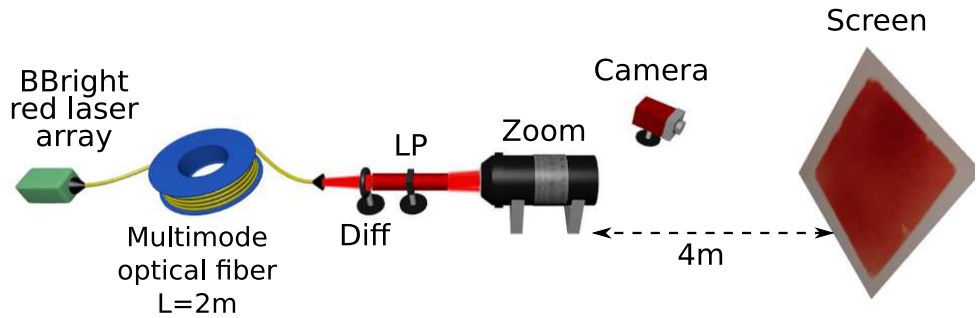}
\caption{{\bf Experimental setup.}
Diff: diffuser, LP: light pipe, Zoom: cinema projector zoom}
\label{fig3}
\end{figure}

In this configuration, we have modified our experimental setup as it is illustrated on the Fig~\ref{fig3}. We have used a red (635nm) visible laser made by BBright (BB-635FC) and based on array semiconductor laser technology. At the output, the beam is injected inside 1mm core diameter multimode optical fiber with two meter length. At the fiber output, the configuration is close to the configuration 3 presented on the Fig~\ref{fig1}A where we can find a diffuser, a light pipe and a projector zoom. The output of the light pipe is projected on a typical projection screen without perforation at a distance of about four meters from the zoom. We can change the image size or the laser power in order to obtain a luminance of about 12cd/m$^{2}$. This value is measured by a standard control device (QALIF) used in the cinema industry and fabricated by Highlands Technologies Solutions. In order to study how to decrease the subjective speckle, we have modified the configuration at different aspects that we will describe in the following.

\begin{figure}[h!]
   \centering
\includegraphics[scale=0.5]{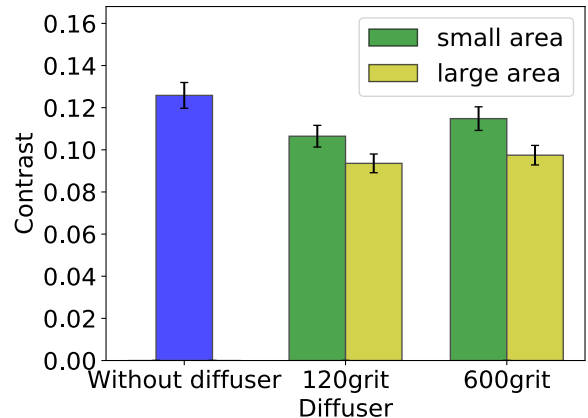}
\caption{{\bf Impact of the incident area illumination on the speckle.}
We have calculated the contrast in a configuration with the light pipe and with or without diffusers. In the last case, we have watched the impact of the incident area illumination of the diffuser on the speckle.}
\label{fig4}
\end{figure}

First, we have measured contrast in a configuration where the diffuser is removed and only the fiber and the light pipe are used. We can see on the Fig.~\ref{fig4} in the blue bar that the contrast is about 12.6\%. However, in this context, the multimode fiber plays already a role to reduce the speckle because of the presence of numerous transversal mode in the fiber (core diameter of one milimeter)~\cite{Goodman:2007,Mehta:2012}. We haven't quantify this quantity because the final result is higher than ten percent. 

Afterwards, we have added either 120 grit or 600 grit diffuser representated on the Fig.~\ref{fig4}. The difference between the green and the yellow bar is the area under which the diffuser is illuminated. In green, we have a small area of illumination because we have placed the fiber very close to the diffuser. In this case, only a small quantity of granularities is illuminated. We obtain a contrast value respectively of about 10.6\% and 11.5\% for 120grit and 600grit. This decrease is small because the phase variation introduce by the diffuser on the incident field is not enough to remove an important part of the speckle. Now, if we increase the illumination area by a factor four, we obtain a contrast value respectively of about 9.35\% and 9.75\% for the 120grit and 600grit. Furthermore, for each barplot, we have calculated the errorbar from the Eq~\ref{eq:delta}. So, first, all the values are out of the incertainty. Then, the incident surface illumination of the diffuser play an important role to reduce the subjective speckle. But the counter part of this operation is important losses of the optical power which has to be compensate by increasing the power of the source if we want to maintain the same screen size.


\vspace{2mm}

\section*{Impact of the magnification on the subjective speckle}
\begin{figure}[!h]
   \centering
\includegraphics[scale=0.5]{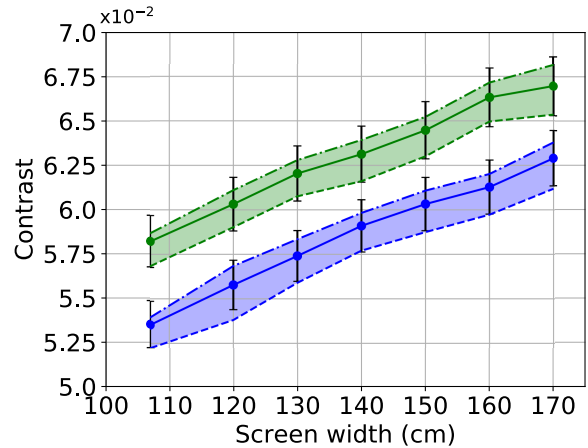}
\caption{{\bf Influence of the objectif magnification on the subjective speckle.}
We have changed the objectif magnification and studied the corresponding contrast for two different diffuser grit. In thick blue line and thick green line, we have respectively the diffuser with 120grit and 600grit. In dashed line, the contrast is calculated from one first part of the image and in points-dashed line for an another part. In black black line, the theoritical error is calculated.}
\label{fig6}
\end{figure}

The last key ingredient in a cinematographic projection system is the simple linear propagation of light towards the screen. In real systems, the output facet of the light pipe is imaged onto the device used to spatially modulate the beam intensity in agreement to the digital content to be displayed. Then, the image of this spatial modulator (nowadays more and more frequently a Digital Light Processor) is projected onto the screen with minimal distortion but a \textit{very large magnification}. Typically, the Digital Light Processor size is of the order of a couple of centimeters, while the screen diagonal is more than ten meters, which implies at least $10^2$ magnification. For the purpose of subjective speckle measurement, the Digital Light Processor is not useful  since aim at measuring deviations from a perfectly flat intensity profile. Therefore, we bypass it and project directly onto the screen the image of the exit facet of the light pipe. On figure \ref{fig6}, we show the measured contrast value for different magnifications, obtained by increasing the propagation distance between the objective and the screen, for two different diffusors. Both curves are shifted with respect to each other due to the different diffusors as expected. Much more importantly the speckle contrast increases with the screen width, \textit{ie} with the magnification after the beam exits the projection system.

\section*{Acknowledgments}

We acknowledge financial support of Région Provence Alpes Côte d'Azur and Bpifrance Financement through contract number DS0025574/00.

\bibliographystyle{bibPi}
\bibliography{article}

\end{document}